\documentclass[twocolumn,conference]{IEEEtran}
\IEEEoverridecommandlockouts

\usepackage{bm}
\usepackage{color}
\usepackage{graphicx}
\usepackage{amsmath}
\usepackage{amssymb}
\usepackage{algorithm}
\usepackage{algorithmic}
\usepackage{ulem}
\usepackage{lineno}
\usepackage{lettrine}
\usepackage{subfigure}
\usepackage{epstopdf}
\usepackage{stfloats}
\usepackage{color}
\usepackage{graphicx}
\usepackage{amsmath}
\usepackage{amssymb}
\usepackage{algorithm}
\usepackage{algorithmic}
\usepackage{amsmath}
\usepackage{multirow}
\usepackage{booktabs}
\usepackage{array}
\usepackage{amsthm}
\usepackage{stfloats}
\usepackage{caption}
\usepackage{subfigure}
\usepackage{bm}
\usepackage{booktabs}
\usepackage{setspace}
\usepackage{makecell}

\newcommand{\be}{\begin{equation}}
\newcommand{\ee}{\end{equation}}
\newcommand{\bea}{\begin{eqnarray}}
\newcommand{\eea}{\end{eqnarray}}
\newcommand{\ba}{\begin{array}}
\newcommand{\ea}{\end{array}}
\newcommand{\nid}{\noindent}

\newcommand{\RNum}[1]{\uppercase\expandafter{\romannumeral #1\relax}}

\allowdisplaybreaks[4]

\title{Joint Beamforming Design in DFRC Systems for Wideband Sensing and OFDM Communications}
\author{ \IEEEauthorblockN{Zichao Xiao$^{\dag}$, Rang Liu$^{\dag}$, Ming Li$^{\dag}$, Yang Liu$^{\dag}$, and Qian Liu$^{\ddag}$
\vspace{-0.0 cm} }
\IEEEauthorblockA{$^{\dag}$ School of Information and Communication Engineering\\Dalian University of Technology, Dalian, Liaoning 116024, China \\ E-mail: \texttt{\{xiaozichao,liurang\}@mail.dlut.edu.cn, \{mli,yangliu\_613\}@dlut.edu.cn} }

\IEEEauthorblockA{$^{\ddag}$ School of Computer Science and Technology \\  Dalian University of Technology, Dalian, Liaoning 116024, China \\ E-mail: \texttt{qianliu@dlut.edu.cn}  \\\; }}
\begin{document}
\maketitle
\pagestyle{empty}
\begin{abstract}
Dual-function radar-communication (DFRC) systems, which can efficiently utilize the congested spectrum and costly hardware resources by employing one common waveform for both sensing and communication (S\&C), have attracted increasing attention.
While the orthogonal frequency division multiplexing (OFDM) technique has been widely adopted to support high-quality communications, it also has great potentials of improving radar sensing performance and providing flexible S\&C.
In this paper, we propose to jointly design the dual-functional transmit signals occupying several subcarriers to realize multi-user OFDM communications and detect one moving target in the presence of clutter.
Meanwhile, the signals in other frequency subcarriers can be optimized in a similar way to perform other tasks.
The transmit beamforming and receive filter are jointly optimized to maximize the radar output signal-to-interference-plus-noise ratio (SINR), while satisfying the communication SINR requirement and the power budget.
An majorization minimization (MM) method based algorithm is developed to solve the resulting non-convex optimization problem.
Numerical results reveal the significant wideband sensing gain brought by jointly designing the transmit signals in different subcarriers, and demonstrate the advantages of our proposed scheme and the effectiveness of the developed algorithm.

\end{abstract}


\section{Introduction}

Dual-functional radar-communication (DFRC), which enjoys high spectrum/power/hardware efficiency through sharing one common waveform for both sensing and communication (S\&C), is regarded as one key enabling technique for the next-generation wireless networks \cite{J A Zhang tutorial}.
Along with the integration and coordination gains, sophisticated designs for the dual-functional waveform are required to handle the conflict requirements of communication and sensing.
Towards this goal, the multiple-input multiple-output (MIMO) architecture has been widely employed in DFRC systems, and the transmit beamforming designs to exploit the spatial degrees of freedom (DoFs) for achieving better performance trade-offs have attracted growing attentions in recent years \cite{RCC Fan Liu}-\cite{Rang Liu DFRC}.


While most of existing works focused on narrowband systems, the research on exploiting wideband signals to improve both sensing and communication performance has begun \cite{JD OFDM DFRC-M F Keskin}-\cite{JD OFDM DFRC-Jeremy Johnston}.
A time-frequency waveform design problem was investigated in \cite{JD OFDM DFRC-M F Keskin} by optimizing the subcarriers powers in the presence of a low-rate feedback channel for conveying transmit waveform control information.
The authors in \cite{JD OFDM DFRC-Zhaoyi Xu} proposed a subcarrier sharing scheme to efficiently exploit the available bandwidth in the orthogonal frequency division multiplexing (OFDM) DFRC system, in which certain subcarriers are shared for realizing S\&C and the other private subcarriers are reserved for exclusive use.
To achieve better S\&C performance, beamforming designs for the transmit signals in different subcarriers should be considered. The authors in \cite{JD OFDM DFRC-Xiaoyan Hu} investigated the low peak-to-average power ratio (PAPR) waveform design problem. The hybrid beamforming with finite-resolution phase shifters was studied in \cite{JD OFDM DFRC-Ziyang Cheng1}, \cite{JD OFDM DFRC-Ziyang Cheng2}. The authors in \cite{JD OFDM DFRC-Jeremy Johnston} considered the design for wideband broadcast systems under communication error rate and beampattern constraints.
In addition, the deployment of intelligent reflecting surface (IRS) in OFDM-DFRC systems was investigated in \cite{JD OFDM DFRC-Tong Wei} to exploit the passive beamforming gain for better S\&C performance.
Although these works verified the advantages of wideband signals in improving both sensing and communication performance, the adopted radar sensing model is very simplified.
Moreover, simply focusing on the joint design of all subcarriers will cause unaffordable computational complexity with limited performance improvement, which is very inefficient in practical OFDM-DFRC systems.


Motivated by above findings, in this paper we propose a more flexible scheme to perform S\&C in OFDM-DFRC systems.
Specifically, the dual-functional signals occupying several frequency subcarriers are jointly designed for multi-user OFDM communications and detecting one moving target in the presence of widely spread clutter, while the signals in other frequency subcarriers are optimized in a similar way to perform other tasks.
After establishing practical models for both sensing and communication functionalities, we investigate to maximize the radar output signal-to-interference-plus-noise ratio (SINR) as well as satisfying the communication SINR requirement and the power budget by jointly optimizing the transmit beamforming and receive filter.
An efficient algorithm based on the majorization minimization (MM) method is developed to solve the resulting complicated optimization problem.
Finally, numerical results reveal the significant sensing gain brought by wideband signals and the advantages of the proposed scheme associated with the developed algorithm.

\begin{figure}[t]
\centering
\includegraphics[width = 2.4 in]{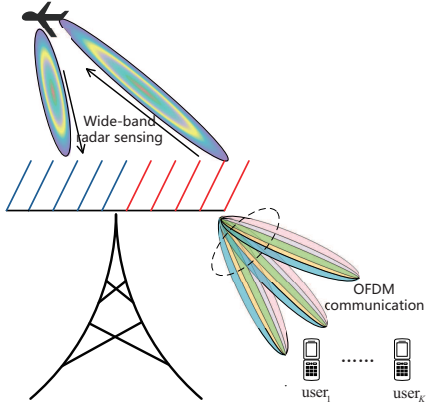}
\caption{The considered wideband OFDM-DFRC system.}\label{system model} \vspace{-1ex}
\end{figure}

\section{System Model and Problem Formulation}

We consider a wideband DFRC system as depicted in Fig. \ref{system model}, where a dual-functional BS equipped with two uniform linear arrays (ULAs) of $N_\mathrm{t}$ transmit antennas and $N_\mathrm{r}$ receive antennas performs S\&C using $N'$ subcarriers.
In particular, the BS occupies several frequency subcarriers of number $N$ to simultaneously serve $K$ single-antenna users and detect one target located at the azimuth direction $\theta_0$ with speed $v_0$ in the presence of widespread clutter.

It is noted that we only use $N\leq N'$ subcarriers to detect one target and communicate with users,
and the residual subcarriers can be flexibly utilized to perform other dual-functional tasks in a similar way.
This scheme is feasible since the orthogonality between different subcarriers naturally supports transferring different data streams to the users without interference.
As for the sensing functionality at the radar receiver, the echoes generated from the considered subcarriers can be easily separated by a bandpass filter, while the interference caused by the adjacent frequency subcarriers with Doppler frequency shifts due to the targets/clutter is negligible.
For example, the echoes in the subcarrier of $f=2.4\mathrm{GHz}$ reflected by a relatively fast target with speed $v=50\mathrm{m}/\mathrm{s}$ induce the Doppler frequency $f^\mathrm{d}=2vf/c=800\mathrm{Hz}$, while a typical OFDM system usually has $N'\geq50$ subcarriers \cite{IEEE standard}, each of which has the bandwidth $\Delta f'\geq 50\mathrm{kHz} \gg f^\mathrm{d}$.
In addition to enabling more flexibilities, the proposed scheme is also necessary in practical applications, since the computational complexity for jointly designing the transmitted signals in different subcarriers will become prohibitively higher with the increase of $N$, while at the same time the performance improvement in target detection introduced by increased subcarriers is very limited.

\subsection{Transmitted Signal Model}

In specific, we denote $\mathbf{s}_{n}[l]\in \mathbb{C}^{K}$, $n=1,\ldots,N$, $l=1,\ldots,L$, as the symbol vector of the $n$-th subcarrier in the $l$-th symbol time-slot for the $K$ users, where $\mathbb{E}\{\mathbf{s}_{n}[l]\mathbf{s}_{n}^H[l]\}=\mathbf{I}_K$, and $L$ denotes the radar pulse/communication frame length.
The symbol vector $\mathbf{s}_{n}[l]$ is passed through a well-designed beamforming matrix $\mathbf{W}_n\triangleq[\mathbf{w}_{n,1},\ldots,\mathbf{w}_{n,K}] \in \mathbb{C}^{N_\mathrm{t}\times K} $, which generates the baseband frequency-domain transmitted signal as
\be
\label{frequency-domain signal i}
\begin{aligned}
\widetilde{\mathbf{x}}_n[l]= \mathbf{W}_n\mathbf{s}_n[l],\quad \forall n.
\end{aligned}
\ee
After $N$-point inverse discrete Fourier transform (IDFT), the cyclic prefix (CP) of size $N_\mathrm{cp}$ is added with duration $T_\mathrm{cp}$ to avoid inter-symbol interference (ISI).
Then, through digital to analog converter (DAC), $\widetilde{\mathbf{x}}_n[l]$, $\forall n$, are transformed into the baseband analog temporal-domain signal as
\be
\label{temporal-domain signal}
{\mathbf{x}}(t)=\sum_{n=1}^N \mathbf{W}_n\mathbf{s}_n[l]e^{j 2\pi (n-1)\Delta ft},
\ee
\nid where $t \in
\big((l-1)(T_\mathrm{s}+T_\mathrm{cp}),
(l-1)(T_\mathrm{s}+T_\mathrm{cp})+ T_\mathrm{s}\big],\forall l$, $\Delta f$ and $T_\mathrm{s}$ are the frequency interval of the OFDM signaling and the OFDM symbol duration, respectively, and we assume $\Delta f=1/T_\mathrm{s}$ to guarantee the orthogonality between different subcarriers.
Finally, the signal is up-converted to the radio frequency (RF) domain via $N_\mathrm{t}$ RF chains with carrier frequency $f_\mathrm{c}$ \cite{JD OFDM DFRC-Ziyang Cheng1} and then emitted through the $N_\mathrm{t}$ antennas.

\vspace{1ex}
\subsection{OFDM Communication Model}

The wideband channel from the BS to user $k$ is modeled by a $D$-tap ($D\leq N_{\mathrm{cp}}$) finite-duration impulse response $\{\mathbf{h}_{k,1},\ldots,\mathbf{h}_{k,D}\}$ \cite{Hongyu Li IRS-OFDM}, where $\mathbf{h}_{k,d}\in \mathbb{C}^{N_\mathrm{t} }$, $\forall d$ are assumed to be perfectly known at the BS.
After down-converting, removing the CP, and $N$-point DFT, the received baseband frequency-domain signal for user $k$ in the $n$-th subcarrier is obtained as
\begin{equation}
\label{frequency-domain received signal for user k within i-th tone}
\begin{aligned}
\widetilde{y}_{n,k}[l]&=\widetilde{\mathbf{h}}_{n,k}^H\mathbf{W}_{n}\mathbf{s}_n[l] +\widetilde{z}_{n,k},\\
\end{aligned}
\end{equation}
where $\widetilde{\mathbf{h}}_{n,k} \in \mathbb{C}^{N_\mathrm{t} }$ denotes the frequency-domain communication channel for user $k$, $\widetilde{z}_{n,k}\sim \mathcal{CN} (0, \sigma^2)$ is the additive white Gaussian noise (AWGN) at the $k$-th user. The corresponding SINR of the $k$-th user in the $n$-th subcarrier can be written as\vspace{1ex}
\begin{equation}
\label{communication SINR}
\begin{aligned}
\mathrm{SINR}_{\mathrm{c},n,k}=\frac{|\widetilde{\mathbf{h}}_{n,k}^H\mathbf{w}_{n,k}|^2}{ \sum_{j\neq k} |\widetilde{\mathbf{h}}_{n,k}^H\mathbf{w}_{n,j}|^2+\sigma^2 }, \forall k, \forall n.\\
\end{aligned}
\end{equation}

\vspace{1ex}
\subsection{Wideband Radar Model}

The received RF echo from the target at the received array can be expressed as\vspace{-1.5ex}

\begin{small}
\be
\mathbf{y}_0^{\mathrm{RF}}\hspace{-0.5ex}(t)\hspace{-0.8ex}
=\hspace{-0.8ex}\sum_{n=1}^N\hspace{-0.4ex} \alpha_{0,n} \mathbf{b}(\hspace{-0.3ex}\theta_0,\hspace{-0.5ex}f_n\hspace{-0.3ex})\mathbf{a}^T(\hspace{-0.3ex}\theta_0,\hspace{-0.5ex}f_n\hspace{-0.3ex}) \hspace{-0.3ex}\mathbf{W}_n\hspace{-0.3ex}\mathbf{s}_n\hspace{-0.3ex}[l]\hspace{-0.3ex} e^{j2\pi(\hspace{-0.3ex}f_\mathrm{n}\hspace{-0.3ex}+\hspace{-0.3ex}f_{0,n}^\mathrm{d}\hspace{-0.3ex} )(\hspace{-0.3ex}t -\hspace{-0.3ex}\tau_0\hspace{-0.3ex})  }, \label{radar RF signal}
\ee\end{small}\vspace{-2ex}

\nid where $t-\tau_0 \in\big((l-1)T_\mathrm{s+cp},(l-1)T_\mathrm{s+cp}+T_\mathrm{s}\big], \forall l$, $\alpha_{0,n}\hspace{-0.5ex} $ represents the target reflection coefficient in the $n$-th subcarrier with $\mathbb{E}\{|\alpha_{0,n} |^2\}=\sigma_{0,n}^2$, $f_{n}\triangleq (n-1)\Delta f +f_\mathrm{c}$ denotes the frequency of the $n$-th subcarrier,
the scalar $\tau_0$ is the two-way propagation delay, $f_{0,n}^\mathrm{d}=2v_0f_n /c$ is the target Doppler frequency of the $ n$-th subcarrier with $c$ representing the velocity of light, and $T_\mathrm{s+cp}\triangleq T_\mathrm{s}+ T_\mathrm{cp}$ for simplicity.
$\mathbf{a}(\theta,f)$ and $\mathbf{b}(\theta,f)$ respectively denote the spatial-frequency steering vector of the transmit and receive signals and are defined as
\be
\label{steering vector model}
\begin{aligned}
&\mathbf{a}(\theta,f) \triangleq\big[1,\hspace{-0.5ex} e^{-j 2 \pi  d_\mathrm{t}\sin \theta /\lambda  }, \ldots, e^{-j 2 \pi\left(N_{\mathrm{t}}-1\right)  d_\mathrm{t}\sin \theta /\lambda }\big]^{T}\hspace{-0.5ex},\\
&\mathbf{b}(\theta,f) \triangleq\big[1,\hspace{-0.5ex} e^{-j 2 \pi d_\mathrm{r}\sin \theta  /\lambda}, \ldots, e^{-j 2 \pi\left(N_\mathrm{r}-1\right)d_{\mathrm{r}} \sin \theta  /\lambda  } \big]^{T}\hspace{-0.5ex},
\end{aligned}
\ee
where $d_\mathrm{t}$ and $d_\mathrm{r}$ denote the transmit and receive antenna spacing, respectively, and $\lambda \triangleq c/f$ denotes the wavelength.

After down-converting, the baseband temporal-domain echo from the target is written as\vspace{-2.5ex}

\begin{small}
\be
\label{baseband temporal-domain radar signal}
\mathbf{y}_0(t)
=\sum_{n=1}^N \alpha_{0,n} \mathbf{b}(\theta_0,f_n)\mathbf{a}^T(\theta_0,f_n) \mathbf{W}_n\mathbf{s}_n[l] e^{j2\pi f_{0,n} (t-\tau_0)  },
\ee
\end{small}\vspace{-1.5ex}

\nid where $f_{0,n}\triangleq f_n+f_{0,n}^\mathrm{d}-f_\mathrm{c} $ denotes the frequency of the $n$-th baseband signal of the target echo and we absorb the constant phase terms associated with $\tau_0$ into the target amplitude for simplicity.
By sampling $\mathbf{y}_0(t)$ $N_\mathrm{s}$ times during each OFDM symbol, the baseband digital samples in the $l$-th time slot can
be obtained as \vspace{-2.5ex}

\begin{small}
\be
\label{baseband temporal-domain radar signal samples}
\begin{aligned}
\mathbf{Y}_{\hspace{-0.3ex}0}\hspace{-0.3ex}[l]\hspace{-0.3ex}\hspace{-0.5ex}
&=\hspace{-0.3ex}\hspace{-0.5ex}\hspace{-0.3ex}\sum_{n=1}^N \hspace{-0.5ex}  \alpha_{0\hspace{-0.3ex},\hspace{-0.3ex}n} e^{\hspace{-0.3ex}j2\pi\hspace{-0.3ex} f_{0\hspace{-0.3ex},\hspace{-0.3ex}n}\hspace{-0.5ex}(l-1\!)T_\mathrm{\hspace{-0.3ex}s+cp\hspace{-0.3ex}}  }\mathbf{p}^T\hspace{-0.5ex}(\hspace{-0.3ex}f_{0,n}\hspace{-0.3ex})\hspace{-0.5ex}\hspace{-0.3ex}\otimes\hspace{-0.5ex}\big(\hspace{-0.3ex} \mathbf{b}(\hspace{-0.3ex}\theta_0,\hspace{-0.3ex}f_n\hspace{-0.3ex})\mathbf{a}^T\hspace{-0.5ex}(\hspace{-0.3ex}\theta_0,\hspace{-0.3ex}f_n\hspace{-0.3ex})\hspace{-0.3ex} \mathbf{W}_{\hspace{-0.3ex}n}\hspace{-0.3ex}\mathbf{s}_n[l]\hspace{-0.3ex}\big),
\end{aligned}
\ee
\end{small}\vspace{-2.5ex}

\nid where $\mathbf{p}(f)\hspace{-0.8ex}\triangleq \hspace{-0.8ex}[e^{j2\pi f (\frac{1}{N_\mathrm{s}})T_\mathrm{s}  },\ldots, e^{j2\pi f (\frac{N_\mathrm{s}}{N_\mathrm{s}})T_\mathrm{s}  } ]^T  $.
Since we treat the received wideband echo as a whole for target detection, there is no need to further convert it into the frequency domain through an invertible DFT operation.
Then, the received baseband digital samples during the $L$ time slots are collected as $\mathbf{Y}_0\triangleq\big[\mathbf{Y}_0[1],\ldots,\mathbf{Y}_0[L] \big]$.
For simplicity, we vectorize $\mathbf{Y}_0$ and transform it into a more concise form as
\begin{align}
\mathbf{y}_0
&\triangleq\big[\mathrm{vec}\{\mathbf{Y}_0[1]\}^T,\ldots, \mathrm{vec}\{\mathbf{Y}_0[L]\}^T \big]^T \nonumber\\
&=\sum_{n=1}^N \hspace{-0.5ex} \alpha_{0,n} \overline{\mathbf{X}}_{n}\big(\mathbf{q}(f_{0,n})\hspace{-0.5ex}\otimes \mathbf{p}(f_{0,n})\otimes \mathbf{b}(\theta_0,f_n) \otimes\mathbf{a}(\theta_0,f_n) \big) \nonumber\\
&=\sum_{n=1}^N   \overline{\mathbf{X}}_{n}\mathbf{v}_{0,n} \nonumber\\
&=\overline{\mathbf{X}} \mathbf{v}_0, \label{baseband temporal-domain radar signal Ns*L samples}
\end{align}
where for brevity we define
\begin{subequations}
\begin{align}
\overline{\mathbf{X}}_{n}
&\triangleq \mathrm{BlkDiag}\big\{\mathbf{I}_{N_\mathrm{s}N_\mathrm{r}} \otimes (\mathbf{s}^T_n[1]\mathbf{W}_n^T) \nonumber\\
&~~~~~~~~~~~~~~~~,\ldots,\mathbf{I}_{N_\mathrm{s}N_\mathrm{r}} \otimes (\mathbf{s}_n^T[L]\mathbf{W}_n^T) \big\}, \label{Definition of X_bar n}\\
\mathbf{q}(f)
&\triangleq [1,\ldots, e^{j2\pi f (L-1)T_\mathrm{s+cp}   } ]^T,\\
\mathbf{v}_{0,n}
&\triangleq \alpha_{0,n}\mathbf{q}(f_{0,n})\hspace{-0.5ex}\otimes\hspace{-0.5ex} \mathbf{p}(f_{0,n})\hspace{-0.5ex}\otimes\hspace{-0.5ex} \mathbf{b}(\theta_0,f_n) \otimes\mathbf{a}(\theta_0,f_n)  ,\\
\overline{\mathbf{X}}
&\triangleq [\overline{\mathbf{X}}_{1},\ldots, \overline{\mathbf{X}}_{N}], \label{Definition of X_bar}\\
\mathbf{v}_0
&\triangleq [\mathbf{v}_{0,1}^T,\ldots,\mathbf{v}_{0,N}^T]^T.\label{Definition of v0}
\end{align}
\end{subequations}

In addition to target echoes, the radar receiver simultaneously receives unwanted clutter echoes that widely spread in both the spatial (e.g., azimuth and range) and Doppler dimensions.
Specifically, we assume that the clutter is generated from the range cell under test and $2M$ other adjacent range cells, each of which contains $N_c$ clutter patches randomly distributed in azimuth.
We set the target range bin as the origin of the range coordinates for simplicity \cite{STAP B Tang}.
Similar to (\ref{baseband temporal-domain radar signal samples}), the received echo from the $n_\mathrm{c}$-th clutter source in the $m$-th cell with the azimuth angle $\theta_{m,n_\mathrm{c}}$ and the speed $v_{m,n_\mathrm{c}}$ is expressed as\vspace{-1ex}
\be
\label{baseband clutter signal samples}
\begin{aligned}
\mathbf{Y}_{m,n_\text{c}}[l]
&\hspace{-0.5ex}=\hspace{-1ex}\sum_{n=1}^N   \alpha_{m,n_\mathrm{c},\hspace{-0.3ex}n} e^{\hspace{-0.3ex}j2\pi\hspace{-0.3ex} f_{m,n_\mathrm{c},\hspace{-0.3ex}n}\hspace{-0.5ex}(l-1\!)T_\mathrm{\hspace{-0.3ex}s+cp\hspace{-0.3ex}}  }\mathbf{p}^T(\hspace{-0.3ex}f_{m,n_\mathrm{c},\hspace{-0.3ex}n}\hspace{-0.3ex})\hspace{-0.5ex}\hspace{-0.3ex}\\
&~~\otimes\hspace{-0.5ex}\big(\hspace{-0.3ex} \mathbf{b}(\hspace{-0.3ex}\theta_{m,n_\mathrm{c},\hspace{-0.3ex}n},\hspace{-0.3ex}f_n\hspace{-0.3ex})\mathbf{a}^T(\hspace{-0.3ex}\theta_{m,n_\mathrm{c},\hspace{-0.3ex}n},\hspace{-0.3ex}f_n\hspace{-0.3ex})\hspace{-0.3ex} \mathbf{W}_{\hspace{-0.3ex}n}\hspace{-0.3ex}\mathbf{s}_n[l]\hspace{-0.3ex}\big)\mathbf{J}_m,
\end{aligned}
\ee
where $\alpha_{m,n_\mathrm{c},\hspace{-0.3ex}n}$ represents the reflection coefficient with , $f_{m,n_\mathrm{c}} \triangleq f_n+f_{m,n_\mathrm{c}}^\mathrm{d}-f_\mathrm{c}$, and the Doppler frequency $f_{m,n_\mathrm{c}}^\mathrm{d}\triangleq2v_{m,n_\mathrm{c}} f_n/c$.
In addition, the shift matrix $\mathbf{J}_{m}\in \mathbb{R}^{N_\mathrm{s}\times N_\mathrm{s}}$, $m=-M,\ldots,M$, is defined by
$\mathbf{J}_{m}(i,j)=\Big\{ \begin{array}{cll}
&1, \quad &i-j+m=0\\
&0, \quad &\mathrm{otherwise}
\end{array}.$
Then, we stack the $L$ samples of the received echo from this clutter source by $\mathbf{Y}_{m,n_\text{c}}\triangleq [\mathbf{Y}_{m,n_\text{c}}[1],\ldots,\mathbf{Y}_{m,n_\text{c}}[L]]$ and vectorize it as \vspace{-1ex}
\be
\label{baseband temporal-domain clutter samples vector}
\begin{aligned}
{\mathbf{y}}_{m,n_\mathrm{c}}=\text{vec}\{\mathbf{Y}_{m,n_\text{c}}\} = \overline{\mathbf{J}}_{m}\overline{\mathbf{X}} \mathbf{v}_{m,n_\mathrm{c}}^\mathrm{c},
\end{aligned}
\ee
where we define $\overline{\mathbf{J}}_{m}\triangleq\mathbf{I}_\mathrm{L}\otimes\big( \mathbf{J}_{m}^T\otimes \mathbf{I}_{N_\mathrm{r}}\big)$, $\mathbf{v}_{m,n_\mathrm{c}}^\mathrm{c}\triangleq[\mathbf{v}_{m,n_\mathrm{c},1}^\mathrm{c},\ldots,\mathbf{v}_{m,n_\mathrm{c},N}^\mathrm{c}]$, and $\mathbf{v}_{m,n_\mathrm{c},n}^\mathrm{c}\triangleq \alpha_{m,n_\mathrm{c},n}\mathbf{q}(f_{m,n_\mathrm{c}})\otimes \mathbf{p}(f_{m,n_\mathrm{c}}) \otimes \mathbf{b}(\theta_{m,n_\mathrm{c}},f_n) \otimes\mathbf{a}(\theta_{m,n_\mathrm{c}},f_n)$.
Therefore, the overall clutter can be written as \vspace{-1ex}
\be
\label{clutter signal Ns*L samples of all clutters}
\begin{aligned}
\mathbf{y}_\mathrm{c}=\sum_{m=-M}^{M}\sum_{n_\mathrm{c}=1}^{N_\mathrm{c}}\mathbf{y}_{m,n_\mathrm{c}}.
\end{aligned}
\ee
The clutter covariance matrix (CCM) is calculated as \vspace{-1ex}
\begin{equation}
\label{clutter covariance matrix}
\begin{aligned}
\mathbf{R}_\mathrm{c}=
\mathbb{E}\{\mathbf{y}_c \mathbf{y}_\mathrm{c}^H\}
=\hspace{-2.5ex}\sum_{m=-M}^{M}
\overline{\mathbf{J}}_{m}\overline{\mathbf{X}}
\mathbf{M}_{m}
\overline{\mathbf{X}}^H\overline{\mathbf{J}}_{m}^H,
\end{aligned}
\end{equation}
where $\mathbf{M}_{m}\triangleq\mathbb{E}\{\sum_{n_\mathrm{c}=1}^{N_\mathrm{c}}
\mathbf{v}_{m,n_\mathrm{c}}^\mathrm{c}(\mathbf{v}_{m,n_\mathrm{c}}^\mathrm{c})^H  \}$ is the inner CCM of the $m$-th range cell.
Instead of requiring the instantaneous information of $\mathbf{v}_{m,n_\mathrm{c}}^\mathrm{c}$, $\forall m$, $\forall n_\mathrm{c}$ for beamforming design, we consider a more practical assumption that only the inner CCMs $\mathbf{M}_{m}$, $\forall m$, are known through prior estimation \cite{CCM estimation}.

To achieve better target detection performance, one linear spatial-temporal receive filter $\mathbf{w}_\mathrm{r}\in \mathbb{C}^{L N_\mathrm{s}N_\mathrm{r}  }$ is employed to process the received echoes.
The output is expressed as 
\begin{equation}
\label{radar output within filter}
\begin{aligned}
y   &=\mathbf{w}_\mathrm{r}^H (\mathbf{y}_0+\mathbf{y}_\mathrm{c}+\mathbf{z}),\\
\end{aligned}
\end{equation}
where $\mathbf{z}\sim \mathcal{CN} (\mathbf{0}, \sigma_\mathrm{r}^2 \mathbf{I}_{L N_\mathrm{s}N_\mathrm{r} })$ denotes the AWGN at the radar receiver. Therefore, the radar output SINR can be written as
\begin{equation}
\label{output radar SINR within filter}
\begin{aligned}
\mathrm{SINR}_\mathrm{r}\hspace{-0.5ex}
        &\triangleq\hspace{-0.5ex}\frac{   \mathbf{w}_\mathrm{r}^H\overline{\mathbf{X}}\mathbf{v}_0\mathbf{v}_0^H\overline{\mathbf{X}}^H\mathbf{w}_\mathrm{r}   }
        { \mathbf{w}_\mathrm{r}^H \left[\sum_{m=-M}^{M}
        \overline{\mathbf{J}}_{m}\overline{\mathbf{X}} \mathbf{M}_{m} \overline{\mathbf{X}}^H\overline{\mathbf{J}}_{m}^H +\sigma_\mathrm{r}^2\mathbf{I}\right] \mathbf{w}_\mathrm{r} }.\\
\end{aligned}
\end{equation}

\subsection{Problem Formulation}

Since the target detection probability is positively related to the radar output SINR, in this paper we aim to maximize $\mathrm{SINR}_\mathrm{r}$ while meeting the communication SINR requirement and the power budget by jointly designing the transmit beamforming $\mathbf{W}_n$, $\forall n$, and the receive filter $\mathbf{w}_\mathrm{r}$.
The optimization problem is thus formulated as
\begin{subequations}
\label{formulated problem}
\begin{align}
\max\limits_{\mathbf{W}_n,\forall n, \mathbf{w}_\mathrm{r}}  ~~& \mathrm{SINR}_\mathrm{r} \label{formulated problem objective function}\\
\text{s.t.} ~~
        & \mathrm{SINR}_{\mathrm{c},n,k} \geq \Gamma_\mathrm{c},\quad \forall k,\forall n,\label{Qos Requirement}\\
        & \sum_{l=1}^L\| \mathbf{W}_n\mathbf{s}_n[l]\|_2^2\leq P_{\mathrm{t},n}, \quad \forall n,
\end{align}
\end{subequations}
where $\Gamma_\mathrm{c}$ is the communication SINR requirement and $P_{\mathrm{t},n}$ is the power budget for the transmitted signals in the $n$-th subcarrier.
Due to the non-convex objective function expressed in (\ref{output radar SINR within filter}) and the complicated fractional terms in (\ref{formulated problem objective function}), (\ref{Qos Requirement}), it is very difficult to directly obtain the solution to problem (\ref{formulated problem}). In the next section, we propose an efficient algorithm based on some sophisticated derivations and the MM method to convert problem (\ref{formulated problem}) into a sequence of favorable sub-problems and iteratively solve them.

\section{Joint Transmit Beamforming and Receive Filter Design}

\subsection{Problem Reformulation}

In order to efficiently solve problem (\ref{formulated problem}), in this subsection we propose to re-formulate it as a more favorable form with less optimization variables.
It is clear that there is no constraint on the radar receiver filter $\mathbf{w}_\mathrm{r}$, i.e., with fixed transmit beamforming $\mathbf{W}_n$, $\forall n$, the optimization problem for $\mathbf{w}_\mathrm{r}$ can be written as an unconstrained problem:
\begin{equation}
\label{problem of u}
\begin{aligned}
\max\limits_{\mathbf{w}_\mathrm{r}}  ~&\frac{   \mathbf{w}_\mathrm{r}^H\overline{\mathbf{X}}\mathbf{v}_0\mathbf{v}_0^H\overline{\mathbf{X}}^H\mathbf{w}_\mathrm{r}   }
        { \mathbf{w}_\mathrm{r}^H \left[\sum_{m=-M}^{M}
        \overline{\mathbf{J}}_{m}\overline{\mathbf{X}} \mathbf{M}_{m} \overline{\mathbf{X}}^H\overline{\mathbf{J}}_{m}^H +\sigma_\mathrm{r}^2\mathbf{I}\right] \mathbf{w}_\mathrm{r} }.\\
\end{aligned}
\end{equation}
This is a well-known generalized Rayleigh quotient whose optimal solution $\mathbf{w}_\mathrm{r}^{\star}$ can be easily obtained as\vspace{-2.5ex}

\begin{small}
\begin{equation}
\label{optimal filter}
\begin{aligned}
\mathbf{w}_\mathrm{r}^{\star}=\frac{   \left[\sum_{m=-M}^{M}
        \overline{\mathbf{J}}_{m}\overline{\mathbf{X}} \mathbf{M}_{m} \overline{\mathbf{X}}^H\overline{\mathbf{J}}_{m}^H +\sigma_\mathrm{r}^2\mathbf{I}\right]^{-1}\overline{\mathbf{X}}\mathbf{v}_0   }
        { \mathbf{v}_0^H\overline{\mathbf{X}}^H \left[\sum_{m=-M}^{M}
        \overline{\mathbf{J}}_{m}\overline{\mathbf{X}} \mathbf{M}_{m} \overline{\mathbf{X}}^H\overline{\mathbf{J}}_{m}^H +\sigma_\mathrm{r}^2\mathbf{I}\right]^{-1} \overline{\mathbf{X}}\mathbf{v}_0 }.
\end{aligned}
\end{equation}
\end{small}\vspace{-2.5ex}

\nid By substituting $\mathbf{w}_\mathrm{r}^{\star}$ into the original problem (\ref{formulated problem}), the joint transmit beamforming and receive filter design problem is reduced to the beamforming design problem as\vspace{-2.5ex}

\begin{small}
\begin{subequations}\label{simplified formulated problem}
\begin{align}
\min\limits_{\mathbf{W}_n,\forall n}  & -{ \mathbf{v}_0^H\overline{\mathbf{X}}^H \Big[\sum_{m=-M}^{M}
        \overline{\mathbf{J}}_{m}\overline{\mathbf{X}} \mathbf{M}_{m} \overline{\mathbf{X}}^H\overline{\mathbf{J}}_{m}^H +\sigma_\mathrm{r}^2\mathbf{I}\Big]^{-1} \overline{\mathbf{X}}\mathbf{v}_0 } \label{objective function of simplified formulated problem} \\
\text{s.t.} ~~
        & \mathrm{SINR}_{\mathrm{c},n,k} \geq \Gamma_\mathrm{c},\quad \forall k,\forall n, \\
        &  \sum_{l=1}^L\| \mathbf{W}_n\mathbf{s}_n[l]\|_2^2\leq P_{\mathrm{t},n}, \quad \forall n.
\end{align}
\end{subequations}
\end{small}\vspace{-2.5ex}

\nid We observe that the objective function is formulated with respect to the matrix $\overline{\mathbf{X}}$, which implicitly contains the optimizing variables $\mathbf{W}_n$, $\forall n$.
To facilitate the following algorithm development, it is necessary to equivalently convert (\ref{objective function of simplified formulated problem}) into an explicit expression with respect to the transmit beamforming $\mathbf{W}_n$, $\forall n$.

Using the definitions in (\ref{Definition of X_bar}) and (\ref{Definition of v0}), the term $\overline{\mathbf{X}}\mathbf{v}_{0}$ in the objective function (\ref{objective function of simplified formulated problem}) can be re-written as
\begin{equation} \label{Xv}
\overline{\mathbf{X}}\mathbf{v}_{0}=\sum_{n=1}^N \overline{\mathbf{X}}_n\mathbf{v}_{0,n}.
\end{equation}
Based on the definition of $\overline{\mathbf{X}}_n$ in (\ref{Definition of X_bar n}), the term $\overline{\mathbf{X}}_n\mathbf{v}_{0,n}$ is re-arranged as \vspace{-1ex}
\begin{equation} \label{Xv expression1}
\overline{\mathbf{X}}_n\mathbf{v}_{0,n}=
\left[\begin{split}
\begin{aligned}
&(\mathbf{I}_{N_\mathrm{s}N_\mathrm{r}} \otimes (\mathbf{s}_n^T[1]\mathbf{W}_n^T)) \mathbf{v}_{0,n,1} \\
&(\mathbf{I}_{N_\mathrm{s}N_\mathrm{r}} \otimes (\mathbf{s}_n^T[2]\mathbf{W}_n^T)) \mathbf{v}_{0,n,2}\\
&~~~~~~~~~~~~~~\vdots\\
&(\mathbf{I}_{N_\mathrm{s}N_\mathrm{r}} \otimes (\mathbf{s}_n^T[L]\mathbf{W}_n^T)) \mathbf{v}_{0,n,L}
\end{aligned}
\end{split}\right],
\end{equation}
where $\mathbf{v}_{0,n}\triangleq[\mathbf{v}_{0,n,1}^T,\ldots, \mathbf{v}_{0,n,L}^T]^T$ with the $l$-th subvector $\mathbf{v}_{0,n,l}\in \mathbb{C}^{N_\mathrm{s}N_\mathrm{r} N_\mathrm{t} }$.
According to the properties of the Kronecker product, the $l$-th term in (\ref{Xv expression1}) can be further transformed into\vspace{-2ex}
\begin{subequations}
\begin{align}
(\mathbf{I}_{N_\mathrm{s}N_\mathrm{r}}\hspace{-0.8ex} \otimes\hspace{-0.5ex} (\mathbf{s}_n^T[l]\mathbf{W}_n^T)) \mathbf{v}_{0,n,l}
&=\mathrm{vec}\{ \mathbf{s}_n^T[l]\mathbf{W}_n^T \mathbf{V}_{0,n,l}\} \\
&=\mathbf{V}_{0,n,l}^T\mathbf{W}_n\mathbf{s}_n[l],\label{Transformation VWs}
\end{align}
\end{subequations} \vspace{-4ex}

\nid where $\mathbf{V}_{0,n,l} \in \mathbb{C}^{ N_\mathrm{t} \times N_\mathrm{s}N_\mathrm{r} }$ is a reshaped version of $\mathbf{v}_{0,n,l}$, i.e., $\mathbf{v}_{0,n,l}=\mathrm{vec}\{\mathbf{V}_{0,n,l}\}$.
Plugging (\ref{Transformation VWs}) into (\ref{Xv expression1}), we have\vspace{-1ex}
\be \label{Transformation Xnvn}
\begin{aligned}
\overline{\mathbf{X}}_n\mathbf{v}_{0,n}&=
\left[\begin{split}
\begin{aligned}
&\mathbf{V}_{0,n,1}^T\mathbf{W}_n\mathbf{s}_n[1] \\
&\mathbf{V}_{0,n,2}^T\mathbf{W}_n\mathbf{s}_n[2]\\
&~~~~~~~~\vdots\\
&\mathbf{V}_{0,n,L}^T\mathbf{W}_n\mathbf{s}_n[L]
\end{aligned}
\end{split}\right]
=\overline{\mathbf{V}}_{0,n}^T\left[\begin{split}
\begin{aligned}
&\mathbf{W}_n\mathbf{s}_n[1] \\
&\mathbf{W}_n\mathbf{s}_n[2]\\
&~~~~~~~~\vdots\\
&\mathbf{W}_n\mathbf{s}_n[L]
\end{aligned}
\end{split}\right]\\
&=\overline{\mathbf{V}}_{0,n}^T (\mathbf{S}_n^T \otimes \mathbf{I}_{N_\mathrm{t}}) \mathrm{vec}\{\mathbf{W}_n\},
\end{aligned}\ee
where $\overline{\mathbf{V}}_{0,n} \triangleq \mathrm{BlkDiag}\{\mathbf{V}_{0,n,1},\ldots,\mathbf{V}_{0,n,L}\}$, $\mathbf{S}_n\triangleq [\mathbf{s}_n[1],\ldots,\mathbf{s}_n[L]]$.
Thus, substituting (\ref{Transformation Xnvn}) into (\ref{Xv}) and replacing the sum operation with equivalent matrix multiplication, we can finally re-write the term $\overline{\mathbf{X}}\mathbf{v}_{0}$ as \vspace{-1ex}
\begin{equation}
\label{Xv transformation}
\begin{aligned}
\overline{\mathbf{X}}\mathbf{v}_{0}
&=\overline{\mathbf{T}}_0\mathbf{w},
\end{aligned}
\end{equation}\vspace{-3ex}

\nid where $\overline{\mathbf{T}}_0\triangleq [\overline{\mathbf{V}}_{0,1}^T (\mathbf{S}_1^T \otimes \mathbf{I}_{N_\mathrm{t}}) ,\ldots, \overline{\mathbf{V}}_{0,N}^T (\mathbf{S}_N^T \otimes \mathbf{I}_{N_\mathrm{t}})]$, and $\mathbf{w}\triangleq[\mathrm{vec}\{\mathbf{W}_1\}^T,\ldots,\mathrm{vec}\{\mathbf{W}_N\}^T]^T$ is a vector form of the transmit beamforming matrices to be optimized.

In order to exploit similar derivations for (\ref{Xv transformation}) to handle the term $\overline{\mathbf{J}}_{m}\overline{\mathbf{X}} \mathbf{M}_{m} \overline{\mathbf{X}}^H\overline{\mathbf{J}}_{m}^H$ in (\ref{objective function of simplified formulated problem}), we first utilize the eigenvalue decomposition to split the semi-definite matrix inner CCMs $\mathbf{M}_{m}$, $\forall m$ as
\be
\begin{aligned}
\mathbf{M}_{m}= \sum_{r=1}^{\mathrm{rank}(\mathbf{M}_{m}) } \!\! \gamma_{m,r} \widetilde{\mathbf{u}}_{m,r}\widetilde{\mathbf{u}}_{m,r}^H =\sum_{r=1}^{\mathrm{rank}(\mathbf{M}_{m}) }  {\mathbf{u}}_{m,r}{\mathbf{u}}_{m,r}^H,
\end{aligned}
\ee
where $\gamma_{m,r}$ denotes the non-zero eigenvalue of $\mathbf{M}_{m}$ with the corresponding eigenvector $\widetilde{\mathbf{u}}_{m,r}$, and we define ${\mathbf{u}}_{m,r}\triangleq \sqrt{\gamma_{m,r}} \widetilde{\mathbf{u}}_{m,r}$ for simplicity.
Then, following the transformation procedures in (\ref{Xv})-(\ref{Xv transformation}), the term $\overline{\mathbf{J}}_{m}\overline{\mathbf{X}} \mathbf{M}_{m} \overline{\mathbf{X}}^H\overline{\mathbf{J}}_{m}^H$ can be equivalently expressed with respect to $\mathbf{w}$ as\vspace{-1ex}
\begin{subequations}
\label{JXu transformation}
\begin{align}
\overline{\mathbf{J}}_{m}\overline{\mathbf{X}} \mathbf{M}_{m} \overline{\mathbf{X}}^H\overline{\mathbf{J}}_{m}^H
&= \sum_{r=1}^{\mathrm{rank}(\mathbf{M}_{m}) }  \overline{\mathbf{J}}_{m}\overline{\mathbf{X}} {\mathbf{u}}_{m,r}{\mathbf{u}}_{m,r}^H \overline{\mathbf{X}}^H\overline{\mathbf{J}}_{m}^H\\
&=\sum_{r=1}^{\mathrm{rank}(\mathbf{M}_{m}) } \overline{\mathbf{T}}_{m,r}\mathbf{w}\mathbf{w}^H \overline{\mathbf{T}}_{m,r}^H, \label{JXu transformation TwwT}
\end{align}
\end{subequations}\vspace{-1.5ex}

\nid where we define $\overline{\mathbf{T}}_{m,r} \triangleq \overline{\mathbf{J}}_{m}[\overline{\mathbf{U}}_{m,r,1}^T (\mathbf{S}_1^T \otimes \mathbf{I}_{N_\mathrm{t}}),\ldots,  \overline{\mathbf{U}}_{m,r,N}^T$\\$ (\mathbf{S}_N^T \otimes \mathbf{I}_{N_\mathrm{t}})]$,
$\overline{\mathbf{U}}_{m,r,n}\triangleq\mathrm{BlkDiag}\{\mathbf{U}_{m,r,n,1},\ldots,\mathbf{U}_{m,r,n,L} \}$, $\mathbf{U}_{m,r,n,l} \in \mathbb{C}^{ N_\mathrm{t} \times N_\mathrm{s}N_\mathrm{r} }$ is a reshaped version of $\mathbf{u}_{m,r,n,l}\in \mathbb{C}^{N_\mathrm{s}N_\mathrm{r} N_\mathrm{t} }$ which is the $l$-th sub-vector of $\mathbf{u}_{m,r,n}\in \mathbb{C}^{LN_\mathrm{s}N_\mathrm{r} N_\mathrm{t} }$, and $\mathbf{u}_{m,r,n}$ is the $n$-th subvector of $\mathbf{u}_{m,r}\in \mathbb{C}^{NLN_\mathrm{s}N_\mathrm{r} N_\mathrm{t} }$.


Therefore, substituting the results in (\ref{Xv transformation}) and (\ref{JXu transformation TwwT}) into (\ref{simplified formulated problem}a), the beamforming design problem can be explicitly and equivalently re-formulated as
\begin{subequations}
\label{reformulated problem}
\begin{align}
\label{objective function of reformulated problem}
\min\limits_{\mathbf{w}}  &~~ -{ \mathbf{w}^H\overline{\mathbf{T}}_0^H \mathbf{A}^{-1}(\mathbf{w}) \overline{\mathbf{T}}_0\mathbf{w} }\\
\text{s.t.} ~
        & \mathrm{SINR}_{\mathrm{c},n,k} \geq \Gamma_\mathrm{c},\quad \forall k,\forall n, \\
        &  \sum_{l=1}^L\| \mathbf{W}_n\mathbf{s}_n[l]\|_2^2\leq P_{\mathrm{t},n}, ~ \forall n,
\end{align}
\end{subequations}
where $\mathbf{A}(\mathbf{w})\triangleq \sum_{m=-M}^{M}
        \sum_{r=1}^{\mathrm{rank}(\mathbf{M}_{m}) } \overline{\mathbf{T}}_{m,r}\mathbf{w}\mathbf{w}^H \overline{\mathbf{T}}_{m,r}^H +\sigma_\mathrm{r}^2\mathbf{I}$.

\subsection{MM-based Transformation}

The beamforming design problem (\ref{reformulated problem}) is still difficult to solve due to the complicated non-convex objective function (\ref{objective function of reformulated problem}).
In this subsection, we propose to utilize the MM method to construct a sequence of more tractable problems to be optimized until convergence.
Specifically, in the $t$-th iteration, a convex surrogate function is constructed to approximate the objective function (\ref{objective function of reformulated problem}) and serve as an upper-bound that should be minimized in the next iteration.
According to \textbf{lemma 1} in \cite{Rang SLP-STAP}, an upper-bound surrogate function for (\ref{objective function of reformulated problem}) can be obtained by
\begin{equation}
\label{new-formulated surrogate function}
\begin{aligned}
(\ref{objective function of reformulated problem})&\leq  \mathrm{Tr} \{
        \mathbf{A}^{-1}(\mathbf{w}_t)\overline{\mathbf{T}}_0\mathbf{w}_t \mathbf{w}_t^H\overline{\mathbf{T}}_0^H\mathbf{A}^{-1}(\mathbf{w}_t)
        \mathbf{A}(\mathbf{w})
        \}\\
&~~-2\mathfrak{R}\{\mathbf{w}_t^H\overline{\mathbf{T}}_0^H\mathbf{A}^{-1}(\mathbf{w}_t)\overline{\mathbf{T}}_0\mathbf{w} \}+c_1\\
&= \mathbf{w}^H\mathbf{U}_t\mathbf{w}-\mathfrak{R}\{\mathbf{b}_t^H\mathbf{w} \}+c_2,
\end{aligned}
\end{equation}
where $c_1$ and $c_2$ are constant terms that are irrelevant to variable $\mathbf{w}$, $\mathbf{w}_t$ denotes the obtained solution in the $t$-th iteration, and
\vspace{-2.5ex}

\begin{small}
\begin{subequations} \label{Update of Ut bt}
\begin{align}
\mathbf{U}_t&\triangleq \hspace{-2ex}\sum_{m=-M}^{M}\hspace{-2ex}
        \sum_{r=1}^{\mathrm{rank}(\mathbf{M}_{m}) } \hspace{-2.5ex}\overline{\mathbf{T}}_{m,r}^H\mathbf{A}^{-1}(\mathbf{w}_t)\hspace{-0.3ex} \overline{\mathbf{T}}_0\mathbf{w}_t \hspace{-0.3ex} \mathbf{w}_t^H\overline{\mathbf{T}}_0^H\mathbf{A}^{-1}(\mathbf{w}_t)\hspace{-0.3ex} \overline{\mathbf{T}}_{m,r}\hspace{-0.3ex} ,\\
\mathbf{b}_t&\triangleq2\overline{\mathbf{T}}_0^H\mathbf{A}^{-1}(\mathbf{w}_t)\overline{\mathbf{T}}_0\mathbf{w}_t.
\end{align}
\end{subequations}
\end{small}\vspace{-2.5ex}

\nid With the above derivations, the optimization problem in each iteration is formulated as
\begin{subequations}
\label{surrogate problem}
\begin{align}
\min\limits_{\mathbf{w}}  ~& \mathbf{w}^H\mathbf{U}_t\mathbf{w}-\mathfrak{R}\{\mathbf{b}_t^H\mathbf{w} \}\\
\text{s.t.} ~
        & \mathrm{SINR}_{\mathrm{c},n,k} \geq \Gamma_\mathrm{c},~ \forall k,\forall n, \\
        &  \sum_{l=1}^L\| \mathbf{W}_n\mathbf{s}_n[l]\|_2^2\leq P_{\mathrm{t},n}, ~ \forall n.
\end{align}
\end{subequations}
It is obvious that this is a typical second-order cone programming (SOCP) problem, which can be readily solved by various existing algorithms or optimization toolboxes.

\begin{algorithm}[t]\begin{small}
  \caption{Joint Transmit Beamforming and Receive Filter Design Algorithm}
  \label{subAlgorithm for max-min}
  \begin{algorithmic}[1]
    \REQUIRE $\overline{\mathbf{T}}_0$, $\overline{\mathbf{T}}_{m,r}$, $\forall m$, $\forall r$, $P_{\mathrm{t},n}$, $\widetilde{\mathbf{h}}_{n,k}$, $\forall n$, $\forall k$, $\sigma$, $\sigma_\mathrm{r}$, $\Gamma_\mathrm{c}$, $\epsilon$.
    \ENSURE  $\mathbf{w}^{\star}$, $\mathbf{w}_\text{r}^{\star}$.
    \STATE {Initialize $t := 0$, initialize $\mathbf{w}_0$ by solving (\ref{initialization problem}).}
    \REPEAT
    \STATE {Update $\mathbf{U}_t$ and $\mathbf{b}_t$ by (\ref{Update of Ut bt});}
    \STATE {Update $\mathbf{w}_{t+1}$ by solving (\ref{surrogate problem});}
    \STATE {$t := t + 1$;}
    \UNTIL {$ \|\mathbf{w}_{t}-\mathbf{w}_{t-1}\|_2/\|\mathbf{w}_{t-1}\|_2 \leq \epsilon$.}
    \STATE {$\mathbf{w}^{\star} := \mathbf{w}_{t}$.}
    \STATE {Calculate $\mathbf{w}_\text{r}^{\star}$ by (\ref{optimal filter}).}
  \end{algorithmic}\end{small}
\end{algorithm}

\subsection{Summary}

Now the proposed joint transmit beamforming and receive filter design algorithm is straightforward and summarized in Algorithm 1, where $\epsilon$ is a parameter to judge the convergence.
In summary, the transmit beamforming $\mathbf{w}$ is iteratively updated by solving (\ref{surrogate problem}) until convergence and then the receive filter $\mathbf{w}_\mathrm{r}^{\star}$ is calculated by (\ref{optimal filter}).
In addition, to provide more flexibility for maximizing radar SINR and ensure a good convergence, we initialize $\mathbf{W}_n$, $\forall n$ by maximizing the minimum communication SINR while satisfying the power budget as
\begin{subequations}
\label{initialization problem}
\begin{align}
\max\limits_{\mathbf{W}_n}~ &\min_{k} ~\mathrm{SINR}_{\mathrm{c},n,k} \\
\text{s.t.} ~
        &  \sum_{l=1}^L\| \mathbf{W}_n\|_F^2\leq P_{\mathrm{t},n}/L,
\end{align}
\end{subequations}
which is a well-known SINR balancing problem and can be easily solved \cite{SINR balancing}. 

\vspace{1ex}
\section{Simulation Results}

In this section, we provide simulation results to illustrate the performance of our proposed algorithm in a wideband DFRC system.
We assume that the BS equipped with $N_\mathrm{t}=N_\mathrm{r}=4$ antennas with spacing $d_\mathrm{t}= 2 c/f_\mathrm{c}$, $d_\mathrm{t}= 0.5 c/f_\mathrm{c}$, serves the $K=3$ users and detects one target located at the azimuth $\theta_0=0^{\circ}$ with speed $v_0=20\mathrm{m/s}$ and reflection power $\sigma_{0,n}^2=-10\mathrm{dB}$, $\forall n$.
The clutter is reflected from 5 range cells with $N_\mathrm{c}=30$ patches, which are uniformly distributed in the azimuth range $(0^{\circ},360^{\circ}]$ and the velocity range $(0,50]$ m/s with power $\sigma_{\mathrm{c},n}^2=-10\mathrm{dB}$, $\forall n$.
The AWGN power at the users and the radar receive is $\sigma^2=-20\mathrm{dB}$ and $\sigma_\mathrm{r}^2=-10\mathrm{dB}$, respectively.
The carrier frequency is $f_\mathrm{c}=2.4\mathrm{GHz}$, the subcarrier spacing is $\Delta f= 0.2 \mathrm{MHz}$, the OFDM data symbol duration is $T_\mathrm{s}= 5\mu \mathrm{s}$, the CP duration $T_\mathrm{cp}=2\mu \mathrm{s}$, and the frame length is $L=8$.
In addition, the parameter for judging convergence is set as $\epsilon=1e-4$.

\begin{figure}[t]
\centering
\includegraphics[width=3.6 in]{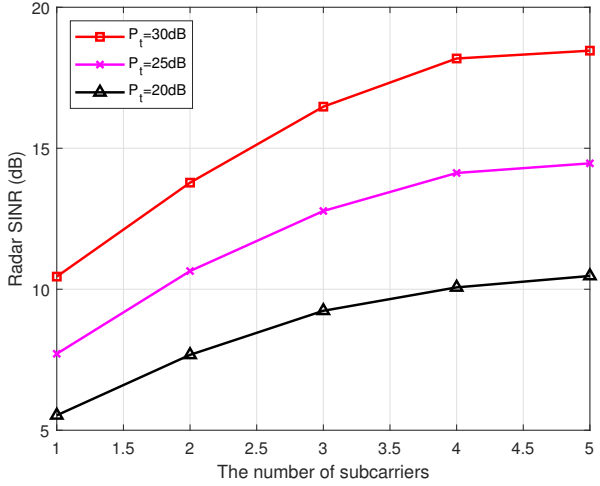}
\caption{Radar SINR versus the number of subcarriers. }\label{simulation1}
\vspace{-1 ex}
\end{figure}

To illustrate the sensing gain brought by the multiple subcarriers in wideband systems, we first plot radar output SINR versus the number of subcarriers in Fig. \ref{simulation1}, where the power budget for each subcarrier is set as $P_{\mathrm{t},n}=P_\mathrm{t}/N$ with $P_\mathrm{t}$ representing the total power for the system, and the number of sampling is $N_\mathrm{s}=5$.
It is obvious that the radar output SINR improves with the increase of the number of subcarriers or the transmit power.
Moreover, we note that the wideband sensing gain provided by adding subcarriers is approaching saturation, e.g., the performance improvement of $N=5$ subcarriers with the total power $P_\mathrm{t}=30\mathrm{dB}$ is only about 0.3dB compared with that of $N= 4$.
This phenomena reveals that jointly designing the transmitted signals in several subcarriers is sufficient to offer satisfactory target detection performance.

\begin{figure}[t]
\centering
\includegraphics[width=3.6 in]{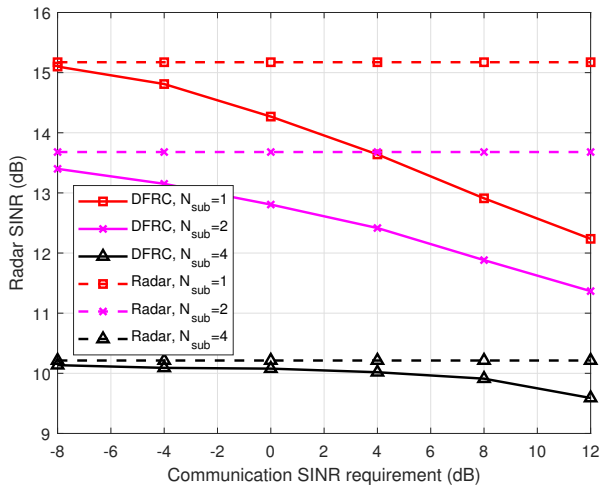}
\caption{Radar SINR versus communication SINR requirement. }\label{simulation2}
\vspace{-1 ex}
\end{figure}

Next, we present the radar output SINR versus the communication SINR requirement in Fig. \ref{simulation2}, where the power budget for each subcarrier is $P_{\mathrm{t},n}=150\mathrm{W}$, $\forall n$, the number of DFRC subcarriers is $N= 4$, the number of delay taps is $D=2$, the $D$-tap response $\mathbf{h}_{k,d}\backsim \mathcal{CN}(\mathbf{0},\mathbf{I}_{N_\mathrm{t}})$, $\forall k$, $\forall d$ and the number of sampling is $N_\mathrm{s}=4$.
For comparisons, we plot three schemes by dividing the $N=4$ subcarriers into $N_\text{sub} = 1, 2, 4$ sets, respectively, and include the scenarios with only radar sensing functionality.
Specifically, the wideband signals in the  $N/N_\text{sub}$ subcarriers of each set are jointly optimized to performance S\&C.
Not surprisingly, the scheme with $N_\text{sub}=1$ achieves the best performance since all the transmitted signals in different subcarriers are jointly designed to fully exploit the wideband sensing gain.
In addition, the trade-off between the communication and sensing performance can be clearly observed.

\section{Conclusions}

In this paper, we investigated the joint beamforming design for a wideband DFRC system, in which several subcarriers are exploited to simultaneously detect one target in the presence of clutter and serve multiple users.
The transmit beamforming and receive filter were jointly optimized to maximize the radar SINR while satisfying the communication SINR requirement and the power budget.
Numerical results demonstrated the remarkable performance improvement brought by wideband signals and revealed the rationality of the proposed scheme.

\end{document}